\documentclass[twocolumn]{aastex631}
\usepackage{amsmath,amstext}
\usepackage[T1]{fontenc}
\usepackage{apjfonts}
\usepackage{graphicx}
\usepackage{placeins}
\usepackage{xspace}

\newcommand{\Aeos}{A{\sc eos}\xspace}
\newcommand{\Msun}{M$_{\odot}$\xspace}

\usepackage{placeins}

\newcommand{\HI}{{H\sc~i}}

\begin{document}

\title{\textsc{Aeos}: The Impact of Population III Initial Mass Function and Star-by-Star Models in Galaxy Simulations}

%% Note that the corresponding author command and emails has to come
%% before everything else. Also place all the emails in the \email
%% command instead of using multiple \email calls.
\correspondingauthor{Kaley Brauer}
\email{kaley.brauer@cfa.harvard.edu}

%\author{}
%\altaffiliation{}
%\affiliation{}
%% The \author command can take an optional ORCID.

\author[0000-0002-8810-858X]{Kaley Brauer}
\affiliation{Center for Astrophysics | Harvard \& Smithsonian, Cambridge, MA 02138, USA}

\author[0009-0006-4744-2350]{Jennifer Mead}
\affiliation{Department of Astronomy, Columbia University, New York, NY 10027, USA}

\author[0000-0003-1173-8847]{John H. Wise}
\affiliation{Center for Relativistic Astrophysics, School of Physics, Georgia Institute of Technology, Atlanta, GA 30332, USA}

\author[0000-0003-2630-9228]{Greg L. Bryan}
\affiliation{Department of Astronomy, Columbia University, New York, NY 10027, USA}

\author[0000-0003-0064-4060]{Mordecai-Mark Mac Low}
\affiliation{Department of Astrophysics, American Museum of Natural History, New York, NY 10024, USA}
\affiliation{Department of Astronomy, Columbia University, New York, NY 10027, USA}

\author[0000-0002-4863-8842]{Alexander P. Ji}
\affiliation{Department of Astronomy \& Astrophysics, University of Chicago, 5640 S Ellis Avenue, Chicago, IL 60637, USA}
\affiliation{Kavli Institute for Cosmological Physics, University of Chicago, Chicago, IL 60637, USA}

\author[0000-0003-2807-328X]{Andrew Emerick}
\affiliation{Carnegie Observatories, Pasadena, CA, 91101, USA}

\author[0000-0003-3479-4606]{Eric P. Andersson}
\affiliation{Department of Astrophysics, American Museum of Natural History, New York, NY 10024, USA}

\author[0000-0002-2139-7145]{Anna Frebel}
\affiliation{Department of Physics and Kavli Institute for Astrophysics and Space Research, Massachusetts Institute of Technology, Cambridge, MA 02139, USA}

\author[0000-0002-9986-8816]{Benoit C{\^o}t{\'e}}
\affiliation{Department of Physics and Astronomy, University of Victoria, Victoria, BC, V8P5C2, Canada}
\affiliation{Konkoly Observatory, Research Centre for Astronomy and Earth Sciences, HUN-REN, Konkoly Thege M. ut 15-17, Budapest 1121, Hungary}

\keywords{Population III stars -- Chemical enrichment -- Dwarf galaxies -- Ionization}

\begin{abstract}
We explore the effect of variations in the Population III (Pop III) initial mass function (IMF) and star-by-star feedback on early galaxy formation and evolution using the \Aeos simulations. We compare simulations with two different Pop III IMFs: $M_\text{char} = 10 \, \mathrm{M}_\odot$ and $M_{\rm max} = 100 \, \mathrm{M}_\odot$ (Aeos10) and $M_\text{char} = 20 \, \mathrm{M}_\odot$ and $M_{\rm max} = 300 \, \mathrm{M}_\odot$ (Aeos20). Aeos20 produces significantly more ionizing photons, ionizing 30\% of the simulation volume by $z \approx 14$, compared to 9\% in Aeos10. This enhanced ionization suppresses galaxy formation on the smallest scales. Differences in Pop III IMF also affect chemical enrichment. Aeos20 produces Population II (Pop II) stars with higher abundances, relative to iron, of light and $\alpha$-elements, a stronger odd-even effect, and a higher frequency of carbon-enhanced metal-poor stars. The abundance scatter between different Pop II galaxies dominates the differences due to Pop III IMF, though, implying a need for a larger sample of Pop II stars to interpret the impact of Pop III IMF on early chemical evolution. We also compare the \Aeos simulations to traditional simulations that use single stellar population particles. We find that star-by-star modeling produces a steeper mass-metallicity relation due to less bursty feedback. These results highlight the strong influence of the Pop III IMF on early galaxy formation and chemical evolution, emphasizing the need to account for IMF uncertainties in simulations and the importance of metal-poor Pop II stellar chemical abundances when studying the first stars.

\end{abstract}

\section{Introduction}

% Hartwig2015: exclude Pop III stars below 0.65 Msun
% Lazar2022: a framework to probe the Pop III IMF by using densities of high-redshift PISNe and GRBs (transients)
% Schneider2006: can explain z~10 sources in UDFs and electron scattering optical depth in WMAP with a standard, slightly top-heavy Larson IMF
% Parsons2022: top-heavy IMF with stars of several hundred solar masses
% GesseyJones2022: variations in 21-cm signal are driven by stars lighter than 20 Msun
% Abel2002: formation of single massive Pop III star in simulation, mass uncertain but >> 1 Msun
% Susa 2014: The mass of the stars found in their simulations are in the range of 1 M ⊙ <~ M <~ 300 M ⊙, peaking at several× 10 M ⊙
% Rossi 2021: using lack of zero-metallicity stars in UFDs to constrain lower mass of Pop III

The formation of the first stars, known as Population III (Pop III), marked a pivotal transition in the evolution of the early Universe \citep{Klessen2023, Barkana2001, Bromm13,Bromm2002,Wise2008}. Composed solely of primordial gas, these stars drove the chemical enrichment of the first galaxies. A key factor in determining the properties of these nascent galaxies is the initial mass function (IMF) of the Pop III stars \citep[e.g.,][]{Schaerer2002,Schneider2006,Hartwig2015,Stacy2016,Ishigaki2018,Lazar2022,Parsons2022}. Understanding the Pop III IMF is crucial for constraining the feedback mechanisms that regulated star formation and the transition to later populations of stars.

Early numerical simulations suggested that the Pop III IMF was top-heavy, favoring very massive stars reaching hundreds or even a thousand solar masses \citep{Abel2002,Bromm2002}. This result stemmed from the inability of the primordial gas to cool efficiently, leading to the formation of large single stars. However, more recent high-resolution simulations have suggested that gas cloud fragmentation could produce a broader range of stellar masses, extending down to tens of solar masses or even lower \citep{Hosokawa11,Hirano14,Susa2014,Stacy2016}. These results suggest that the Pop III IMF (which is still distinct from a classical Pop II IMF) encompasses a broad range of stellar masses, allowing for the formation of both extremely massive and relatively lower-mass stars in the early universe.

Indirect observational evidence further supports the idea of a broad Pop III IMF. Metal-poor stars observed in the Milky Way and its satellite galaxies retain chemical abundance patterns that may represent enrichment by Pop III stars \citep{Hartwig2018}. When these patterns are compared to theoretical models of Pop III supernova yields, they can offer insights into the masses of Pop III progenitors. Some data suggest that certain enriched stars originated from Pop III progenitors with masses exceeding 50 \Msun \citep{Ji24} and even up to 260 \Msun \citep{Xing23}; however, the latter finding is contested by recent work \citep{Skuladottir2024}. Meanwhile, the majority of Pop III progenitors may have had masses of only a few tens of solar masses \citep{Ishigaki2018,Fraser17,Tumlinson2006}. While some models propose a Salpeter-like IMF to describe the Pop III mass distribution, recent studies have pointed to possible deviations from this power law, including a potential peak around 25 \Msun \citep{Ishigaki2018}. The non-detection of zero-metallicity stars in ultra-faint dwarf galaxies also provides constraints on the minimum mass of the first stars, $M_{\text{min}}>0.8$ \Msun \citep{Rossi21}.

Despite these advances, significant uncertainties remain regarding the true shape of the Pop III IMF and its implications for galaxy formation and reionization. Simulations provide a valuable tool for exploring these uncertainties by modeling different IMF scenarios and examining their effects on key cosmic processes. The amount of ionizing radiation produced by Pop III stars, for instance, depends heavily on their mass distribution, directly impacting the ionization history of the universe and star formation in early galaxies \citep{Schaerer2002,WiseAbel2012,Kimm2017}. Likewise, stellar feedback from SNe and winds influences the growth and chemical evolution of the earliest galaxies \citep{Wise2012a,HegerWoosley2010}.

The legacy of the Pop III stars extends to the formation of Population II (Pop II) stars \citep[e.g.,][]{Prgomet22}. The Pop III IMF influences the characteristics of the Pop II stars, which emerged after the interstellar medium was enriched by metals from Pop III SNe. Massive Pop II stars play a critical role in early galaxy evolution as they contribute significantly to the regulation of star formation through feedback processes such as ionizing radiation, stellar winds, and SNe \citep{SomervilleDave2015}.
These feedback mechanisms drive gas outflows, influence the ionization history, and regulate the growth of early galaxies. Accurately capturing the effects of individual Pop II stars on their environment is therefore also essential for understanding the formation and evolution of the first galaxies \citep[e.g.,][]{AeosMethods,Andersson24}. 

Traditional cosmological simulations often simplify the modeling of Pop II stellar feedback by representing entire star clusters as single particles, which limits the resolution of the feedback processes that individual stars provide \citep[e.g.,][]{Skinner20,Smith2021}. Such models can overlook the detailed radiative and mechanical feedback from individual stars, which plays a critical role in regulating the interstellar medium and shaping galaxy evolution. In recent years, simulations have begun to include feedback from individual stars in galaxy simulations \citep[e.g.,][]{Emerick2019,Lahen2020,Hirai21,Gutcke21,Hislop22,Andersson23,Calura22,Steinwandel23,AeosMethods,Hirai24,Deng24}. Star-by-star modeling of Pop II stellar feedback offers a more detailed approach. This captures the impact of the radiation, supernovae (SNe), and stellar winds from individual stars both inside and outside clusters on gas dynamics and chemical enrichment, which is crucial for understanding the earliest stages of galaxy formation.

In this work, we present results from the \Aeos simulations \citep{AeosMethods,Mead2024b} that explore (1) the impact of varying the Pop III IMF on early galaxy evolution and (2) the impact of star-by-star models when simulating the first galaxies. These \Aeos simulations model the formation of the first stars and galaxies in the first few hundred megayears of the universe, following individual stellar feedback and detailed chemical enrichment. This approach is motivated by the fundamental importance of predicting the chemical abundances of ancient stars, particularly those found in ultra-faint dwarf galaxies and the Milky Way's stellar halo, to unravel the processes of early star and galaxy formation \citep[e.g.,][]{Brauer19}.

In Section \ref{sec:methods}, we summarize our methodology, describe our simulations, and detail how they differ. In Section \ref{subsec:Aeos10vs20}, we compare two individual star simulations %, Aeos10 and Aeos20,
that differ only in their Pop III IMFs. These simulations allow us to investigate how changes in the IMF affect star formation rates, galaxy growth, ionization history, and metal enrichment. In Section \ref{subsec:noindiv}, we compare one of the simulations %Aeos10 
to seven comparison simulations without star-by-star feedback to investigate the effect of highly resolved star formation and feedback on galaxy evolution. We conclude in Section \ref{sec:conc}.

\section{Methods} \label{sec:methods}

This study uses the numerical methods detailed in \citet{AeosMethods}. We focus on a comparative analysis between the fiducial A{\sc eos} cosmological simulation and additional simulations: (1) an A{\sc eos} simulation with an altered Pop III IMF with a higher characteristic mass $M_{\rm char}$ and upper mass cutoff, and (2) seven comparison simulations that omit the full star-by-star feedback and enrichment models, instead using single stellar population particles and IMF averaged metal yields. The comparison simulations are identical except for differences arising primarily from stochastic IMF sampling during star formation. We run seven comparison simulations to show how stochasticity affects results. Below, we provide a concise summary of the important methods from \citet{AeosMethods} and elaborate on the modifications pertinent to the additional simulations. The differences in the simulations are summarized in Table \ref{tab:simoverviews}.

\subsection{Similarities Across Simulations}

% \subsubsection{Numerical Framework}

All simulations use the {\sc Enzo} adaptive mesh refinement (AMR) code \citep{Enzo2014,Enzo2019}, enabling high-resolution modeling of cosmological structures. The computational grid employs AMR to achieve high physical resolution in regions of elevated gas density, ensuring detailed resolution of dense gas clumps and star formation. They have a root-grid resolution of $256^3$, a dark matter resolution of 1840 M$_\odot$, and a resolution of 1 physical pc at the finest scales.
%Hydrodynamic equations are solved using the Piecewise Parabolic Method (PPM) \citep{ColellaWoodward1984,Bryan1995}, supplemented by a two-shock approximate Riemann solver to maintain numerical stability. Gravity is computed from gas self-gravity, star particles, and a static dark matter background via a multigrid Poisson solver and an adaptive particle-mesh N-body solver with an effective force resolution of approximately $2\Delta x$, where $\Delta x$ denotes the local cell size.

% \subsubsection{Initial Conditions}

The simulations cover a comoving (1 Mpc)$^3$ volume, simulated from redshift $z = 130$ to $z=14.5$, approximately~300 Myr after the Big Bang. 
They commence from identical initial conditions generated with {\sc MUSIC} \citep{Hahn2011} at redshift $z = 130$, adhering to the cosmological parameters derived from the Planck collaboration \citep{Planck2014}. This ensures consistency across all simulation runs, facilitating comparisons between different feedback models and IMF choices.

% \subsubsection{Star Formation}

Star formation is modeled stochastically in cold, dense gas regions with $n > 10^4$ cm$^{-3}$ that exhibit converging flows having $\nabla \cdot \mathbf{v} < 0$, assuming a star formation efficiency per free-fall time of $e_{\rm ff} = 2\%$. The star formation algorithm distinguishes Pop II from Pop III stars using a threshold of total gas metallicity $Z > 10^{-5} Z_{\odot}$ \citep{Ji2014, Chiaki2015,Schneider2012,Tumlinson2006}. Pop III star formation in lower metallicity gas further requires a molecular hydrogen fraction $f_{\rm H_2} > 0.0005$, consistent with high-resolution simulations \citep{Susa2014,Kulkarni2021}. 

% \subsubsection{IMF} \label{subsec:pop3IMF}

Following \citet{Wise2012a}, we adopt a Pop III IMF of the form:
\begin{equation}
f(\log M) \, \mathrm{d}M = M^{-1.3} \exp\left[ - \left( \frac{M_\text{char}}{M} \right)^{1.6} \right] \, \mathrm{d}M
\end{equation}
Above $M_\text{char}$, it behaves as a Salpeter IMF, and below $M_\text{char}$ it is exponentially suppressed \citep{Chabrier2003}. We also include an upper mass limit $M_{\rm max}$. In all simulations, every Pop III star is represented by a single star particle. The Pop II IMF is sampled from a \citet{Kroupa2001} distribution with a mass range of 0.08--120 \Msun.

%While the Pop III IMF differs between different A{\sc eos} simulations (see Section \ref{subsec:pop3IMF}), 

% \subsubsection{Stellar Feedback}

All simulations include stellar feedback from stellar radiation and SNe: \begin{itemize} \item Stellar Radiation: Radiation in multiple bands (IR, FUV, LW, \HI, He{\sc i}, and He{\sc ii}) is tracked using adaptive ray-tracing radiative transfer methods. \item SNe: Core-collapse (both Pop III and Pop II) and Type Ia SNe inject mass, energy, and metals into the ISM with event-specific yields and energies. \end{itemize}

We follow radiation from every star with mass $M > 8$~\Msun. For Pop III stars, we use binned photon counts from \citet{HegerWoosley2010} with the lifetimes in \citet{Schaerer2002}. For Pop II stars, photon fluxes in each radiation band are determined using the OSTAR2002 \citep{Lanz2003} grid of O-type stellar models. We also use the \textsc{PARSEC} \citep{Bressan2012} grid of stellar evolution tracks to set the lifetime of each Pop II star and the start time and length of the asymptotic giant branch (AGB) phase, if present.

%that is not included in the comparison simulations; see Section \ref{subsec:diff:feedback}.

% \subsubsection{Radiative Cooling and Chemistry}

All simulations employ a modified version of {\sc grackle} \citep{GrackleMethod} to handle non-equilibrium chemistry involving nine species (H, H$^+$, He, He$^+$, He$^{++}$, e$^{-}$, H$^-$, H$_{2}$, and H$_2^+$) and associated radiative cooling and heating processes. The chemical network includes dust-mediated H$_2$ formation, a UV background that extends to high redshift and is scaled to be continuous with \citet{HM2012} at $z=10$, and a Lyman-Werner background model from \citet{Emerick2019} and \citet{Wise2012a} that adopts updated rates at high redshift from \citet{Qin2020}.

\subsection{Differences Between Simulations}

The two A{\sc eos} simulations (Aeos10 and Aeos20) differ in the characteristic mass $M_{\rm char}$ and upper limit $M_{\rm max}$ of their Pop III IMFs. The comparison simulations share the same Pop III IMF as Aeos10, but do not have star-by-star modeling for Pop II stars, do not model feedback from winds, and do not track detailed chemical enrichment with individual metals. These differences are summarized in Table \ref{tab:simoverviews}.

\begin{table*}[htbp]
  \caption{Overview of \Aeos Simulation Runs}
  \label{tab:simoverviews}
  \begin{tabular}{lccccc}
    \hline
    \textbf{Run Name} & \textbf{Pop III $M_{\rm char}$} & \textbf{Pop III $M_{\rm max}$} & \textbf{Pop II Star Resolution} & \textbf{Individual Metals} & \textbf{Stellar Winds} \\
    \hline 
    Aeos10 & 10 \Msun & 100 \Msun & 2 M$_{\odot}$ & Yes & Yes \\ 
    Aeos20 & 20 \Msun & 300 \Msun & 2 M$_{\odot}$ & Yes & Yes\\ 
    Comparisons & 10 \Msun & 100 \Msun  & 1000 M$_{\odot}$ & No & No \\
    \hline
  \end{tabular}
  \hfill
\end{table*}

Aeos10 adopts a Pop III IMF with a characteristic mass $M_{\rm char} = 10$~\Msun and a maximum mass of $M_{\rm max} = 100$~\Msun. To investigate the sensitivity of galaxy evolution to the Pop III IMF, we execute an additional \Aeos simulation (Aeos20) wherein the characteristic mass $M_{\rm char}$ of the Pop III IMF is modified from the fiducial value of 10 \Msun to 20~\Msun and $M_{\rm max} = 300$~\Msun \citep[motivated by][]{Hirano17, Bromm13,Yoshida06,Skinner20}. This adjustment alters the mass distribution of Pop III stars, thereby affecting their lifetimes, feedback outputs, and chemical yields \citep[see][]{AeosMethods}. This is the only difference between the Aeos10 and Aeos20 simulations. 

%\subsubsection{Star Particle Resolution}

In the \Aeos simulations, Aeos10 and Aeos20, star particles represent individual stars sampled from their respective IMFs except for Pop II stars below 2~\Msun, which are aggregated into single particles due to their negligible feedback and enrichment contributions on simulation timescales. Star formation occurs from gas reservoirs of $\geq 100$ \Msun. Individual star particles are formed by sampling from the IMF until the gas reservoir is depleted, at which time star formation ends until more gas meets the conditions for star formation.

In the comparison simulations, Pop III stars are represented by individual particles, but Pop II stars are represented by star cluster particles with a minimum mass of 1000 \Msun. For Pop II star formation, stars are formed from gas reservoirs of $\geq 1000$ \Msun. A single star particle is formed with a stellar mass distribution determined by sampling from the IMF.

%\subsubsection{Stellar Feedback} 
%\label{subsec:diff:feedback}

In addition to feedback from CCSNe and radiation, \Aeos simulations include feedback and enrichment from AGB winds and massive star winds. 
The seven comparison simulations presented here omit AGB and massive stellar winds. They also omit individual star-by-star feedback modeling due to their aggregation of Pop II stars into cluster particles. 

Additionally, the comparison simulations track overall metallicity but do not track individual elements.
The \Aeos simulations track the evolution of ten individual metal abundances -- C, N, O, Na, Mg, Ca, Mn, Fe, Sr, and Ba -- in addition to hydrogen and helium. These elements are sourced from various nucleosynthetic channels, including core-collapse SNe, Type Ia SNe, and asymptotic giant branch stars. We use detailed yield tables from \citet{HegerWoosley2010} for Pop III CCSNe and \citet{Limongi2018} for Pop II stars.

\section{
%Comparison between Aeos10 and Aeos20: 
Effects of Varying Pop III IMF} 

\label{subsec:Aeos10vs20}

\subsection{Ionization}% from Different Pop III IMFs}

\begin{figure}[tb]
\center
\includegraphics[width=\linewidth]{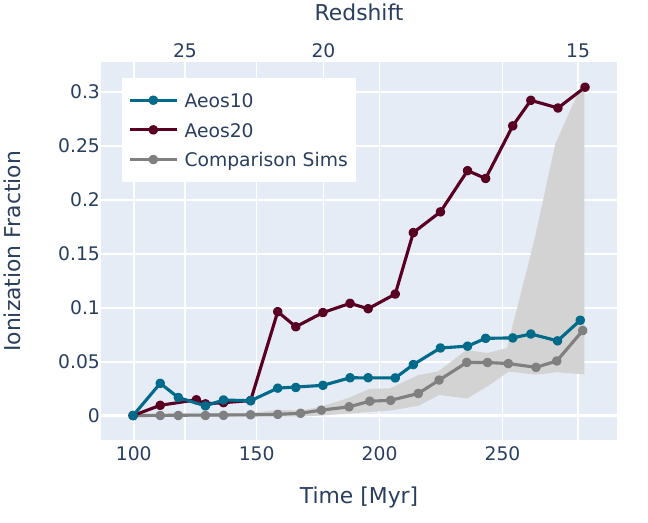}
\caption{Ionization fraction of the simulation volume over time for Aeos10, Aeos20, and the comparison simulations using the Aeos10 Pop III IMF but without individual stellar feedback. The shaded grey region shows the 16th to 84th percentile of ionization fractions for the different comparison simulations. Aeos20 ionizes the volume significantly earlier than Aeos10 due to the more massive Pop III stars producing more ionizing photons. The comparison simulations have similar ionization to Aeos10, as expected, with stochastic differences beginning to cause large variations between simulations after around 250 Myr. The choice of Pop III IMF significantly affects the ionization and consequently the growth of the smallest galaxies.
\label{fig:ionization}}
\end{figure}

The different Pop III IMFs in Aeos10 and Aeos20 lead to a pronounced difference in the hydrogen ionization of the simulation volume. Figure \ref{fig:ionization} shows that Aeos20 ionizes much more rapidly than Aeos10, with the volume 30\% ionized by $z=14.5$ in Aeos20 compared to 9\% in Aeos10.

\begin{figure}[tb]
\center
\includegraphics[width=\linewidth]{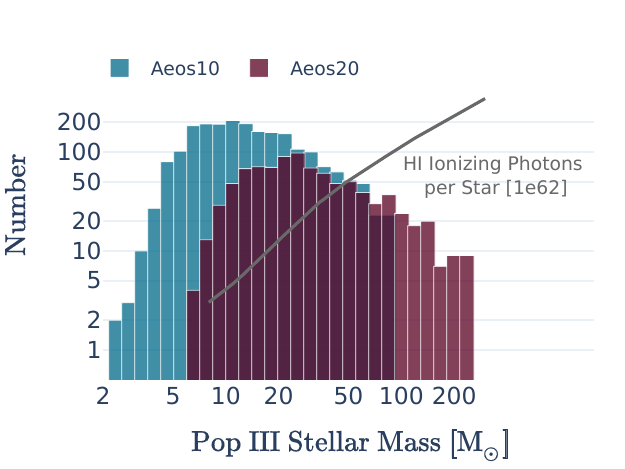}
\caption{Distribution of Pop III star masses in Aeos10 vs. Aeos20. Aeos10 has far more low-mass Pop III stars, while Aeos20 has fewer total Pop III stars but their masses extend up to 300 \Msun. The black line shows the number of HI ionizing photons from each star as a function of stellar mass; several dozen Pop III stars with >~100~\Msun in Aeos20 result in a large number of ionizing photons.
\label{fig:p3_hist}}
\end{figure}

\begin{figure}[t]
\center
\includegraphics[width=0.9\linewidth]{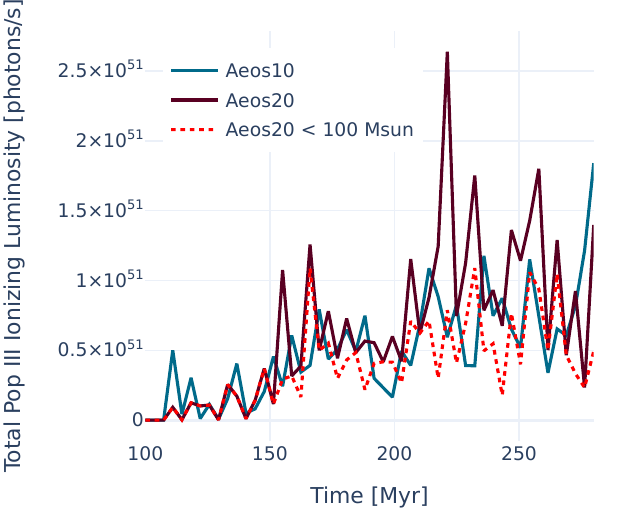}
\caption{Starting after $\sim150$ Myr, Aeos20 (maroon) begins to exceed Aeos10 (blue) in the total ionizing luminosity from Pop III stars. This excess of ionizing photons is primarily due to the luminosity of stars greater than 100 \Msun, because when the contribution of these stars is removed (dotted red), the total luminosity of the two simulations is far more similar.
\label{fig:p3_lum}}
\end{figure}

\begin{figure*}[bt]
\center
\includegraphics[width=\linewidth]{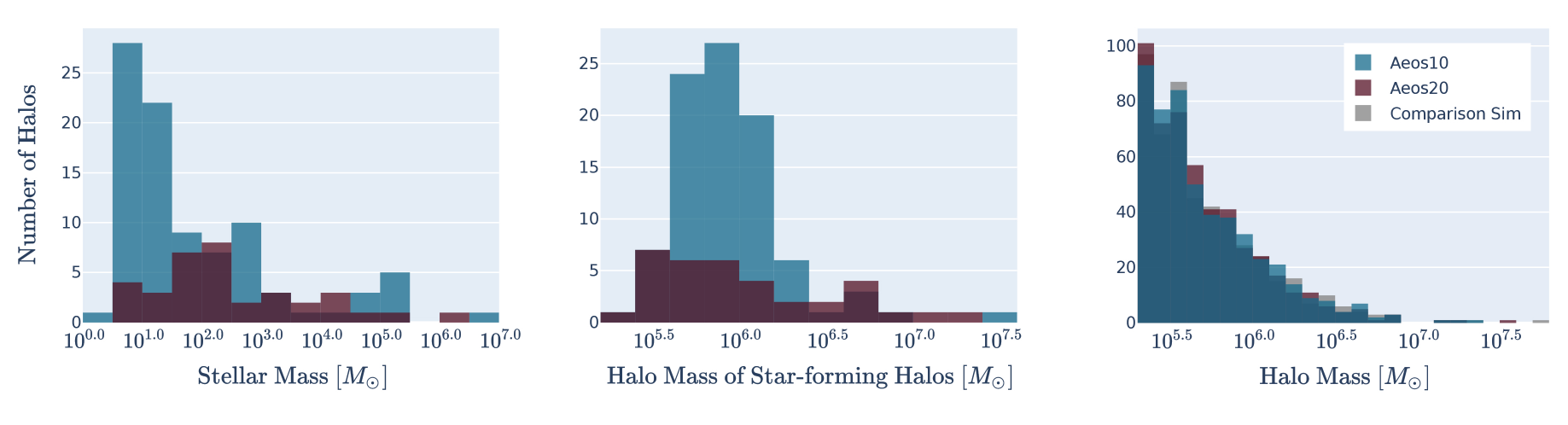}
\caption{Stellar masses (\textit{left}) and halo masses (\textit{middle}) of star-forming halos in the different simulations. Aeos10 has many more small star-forming halos than Aeos20. This is due to higher ionization in Aeos20 (see Figure \ref{fig:ionization}). The distribution of halo masses for all halos (\textit{right}) in all simulations are nearly identical, as expected. \label{fig:haloHists_a20}}
\end{figure*}

\begin{figure}[t]
\center
\includegraphics[width=0.9\linewidth]{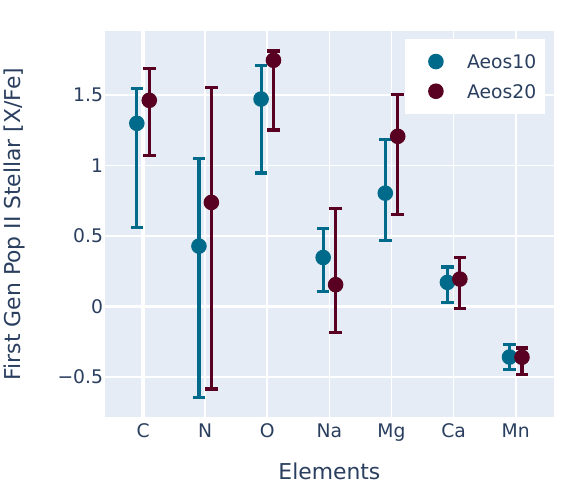}
\caption{Median stellar chemical abundances ([X/Fe]) of the oldest low-mass Pop II stars from each halo that has started Pop II star formation in the simulation with 16th to 84th percentile scatter between the different halos. The mass of Pop III stars affects their yields, affecting the abundances in first generation Pop II stars. In particular, light elements (C, N) and $\alpha$-elements (O, Mg) are increased in the yields of higher-mass Pop III stars, and elements with even atomic numbers (e.g., Mg) are more abundant relative to odd-numbered elements (e.g., Na). The scatter between halos is significant compared to the differences due to adopted Pop III IMF, however.
\label{fig:p2_XH}}
\end{figure}

This occurs because the ionizing radiation emitted by a star increases with its mass \citep{Schaerer2002}. Consequently, the IMF in Aeos20, with its higher $M_\text{char}$ and upper mass limit, produces a stellar population that emits significantly more ionizing photons. Figure \ref{fig:p3_hist} shows the mass distribution of Pop III stars in the two simulations. Aeos10 contains approximately 2000 stars with masses in the tens of solar mass range, while Aeos20 has fewer than 500 stars in this range. However, Aeos20 features Pop III stars up to 300~\Msun, including about 80 stars exceeding 100~\Msun. The total number of ionizing photons from Pop III stars per stellar mass for these two distributions (combining the histograms and rate of ionizing photons shown in Figure \ref{fig:p3_hist}) is about $7 \times 10^{61}$ photons/\Msun for Aeos10 and $10 \times 10^{61}$ photons/\Msun for Aeos20.

These massive stars are the primary contributors to the increased ionizing photon output, as shown in Figure \ref{fig:p3_lum}. The total ionizing luminosity of Aeos20 surpasses Aeos10 due to its most massive stars. When only stars below 100~\Msun are included, the total ionizing luminosity of Aeos20 agrees fairly well with that of Aeos10; the major peaks in ionizing luminosity are no longer seen. Thus, the enhanced ionization in Aeos20 seen in Figure \ref{fig:ionization} is predominantly driven by its population of Pop III stars exceeding 100~\Msun.

The characteristic mass of the IMF also affects the ionization.
If enhanced ionization was driven only by the most massive Pop III stars, we would expect the ionizing output from Aeos20 stars below 100~\Msun to be significantly lower than in Aeos10, as we are excluding a substantial portion of stellar mass. However, since the total ionizing luminosity remains comparable to Aeos10, this indicates that the characteristic mass also contributes to ionizing luminosity, albeit to a lesser extent than the high-mass tail of the IMF. Additionally, in terms of total stellar mass, Aeos20 actually produces somewhat less stellar mass than Aeos10 at both early and late times, emphasizing that the higher ionization is due to the distribution of Pop III star masses rather than differences in total stellar mass.

The higher ionization in Aeos20 suppresses the formation of small galaxies. Previous work has found that ionization-driven photoheating of the gas in and around dwarf galaxies suppresses star formation \citep{Efstathiou1992,Bullock2000}, particularly in halos with $M_\text{halo} \lesssim 10^9$ \Msun \citep[e.g.,][]{Thoul1996,Okamoto2008,Dawoodbhoy2018}. In our simulations, all galaxies exist in small halos and we find they are highly sensitive to the rate of reionization.
As seen in Figure \ref{fig:haloHists_a20}, Aeos10 has far more galaxies at low halo masses. For $M_* < 100$ M$_\odot$, Aeos10 has 60 star-forming halos while Aeos20 has only 14 at $z=14.5$. Meanwhile, the distributions of halo masses for all halos are essentially identical for Aeos10, Aeos20, and the comparison simulations, as expected for simulations with the same dark matter initialization. 
Another mechanism, Lyman-Werner (LW) radiation, can also suppress early star formation by dissociating H$_2$, the primary coolant in metal-poor gas \citep[e.g.,][]{Safranek2012}. Both Aeos10 and Aeos20 have identical LW backgrounds, however, so the suppression of small galaxy formation in Aeos20 vs. Aeos10 is isolated to the differences in ionizing photons. At the beginning of the simulation ($\sim 120$ Myr), Aeos10 briefly surpasses Aeos20 in ionizing photons (see Figures \ref{fig:ionization} and \ref{fig:p3_lum}). This is due to stochastic differences during the beginning of Pop III star formation, when there are very few stars, and a slightly delayed onset of Pop III star formation in Aeos20. An ensemble of simulations with Aeos10 and Aeos20 parameters would be preferred to better understand stochastic differences, but is currently unfeasible due to the computational cost of the simulations.

We selected Aeos10 as our fiducial simulation in \citet{AeosMethods} because the high level of ionization in Aeos20 causes reionization to occur much earlier than expected: 30\% ionized at redshift $z \sim 14$ compared to the inferred redshift of $z \sim 8$ from \citet{planck2016}. 
Our simulation volume is a small region that is not generally representative of the very early Universe, however. We thus avoid interpreting this early reionization as a constraint on the Pop III IMF, which remains highly uncertain. Instead, we interpret the differences between Aeos10 and Aeos20 as evidence of the strong impact that different Pop III IMF choices have on reionization and the number of small, early galaxies. Given both the importance of the Pop III IMF and its uncertainty, simulations of the first galaxies must take care to understand the impact of Pop III IMF choices on their results.

\subsection{Chemical Enrichment}% from Different Pop III IMFs}

Nucleosynthetic yields from Pop III core-collapse SNe depend on their masses \citep{HegerWoosley2010}. These yields differ not only in total metal production but also in the relative abundances of individual metals \citep{Rossi2024,AeosMethods}.

When we vary the Pop III IMF, we thus produce different chemical abundances in the subsequent Pop II stars. In Figure \ref{fig:p2_XH}, we show how the differences in yields translate into the chemical abundances of the oldest Pop II stars. We show the median stellar chemical abundances of first-generation Pop II stars from each galaxy in the simulations that has begun Pop II star formation (17 galaxies in Aeos10, 15 galaxies in Aeos20). We limit the analysis to low-mass stars, e.g., the $\leq 0.8$ M$_\odot$ stars that would survive to present day in the unresolved star particles. 

While stellar chemical abundance differences are apparent in the Figure, we note that the halo-to-halo scatter is significant for some of the elements as shown by the scatter bars. To quantify the number of Pop II halos required to distinguish differences resulting from these Pop III IMFs, we estimate the sample size necessary for the mean abundances to separate beyond the intrinsic halo-to-halo scatter. For elements such as C and N, more than 20–30 halos are needed to achieve statistical significance, while elements such as Mg and O require fewer than 10. Given the current sample of 15-17 Pop II halos per simulation, this suggests that stochastic enrichment dominates over systematic trends for several elements, emphasizing the need for larger sample sizes when interpreting the impact of Pop III IMF variations on early chemical enrichment.

Pop II stars in Aeos20 exhibit higher [X/Fe] values for light elements such as nitrogen (N) and oxygen (O), reflecting the increased nucleosynthetic yields from more massive Pop III stars due to stronger $\alpha$-capture reactions and rotational mixing \citep{HegerWoosley2010}.
Carbon (C) is also enhanced in Aeos20, primarily due to the significant production of carbon during helium burning in more massive Pop III stars. In these simulations we do not include pair-instability SNe \citep{Rakavy1967,Fraley1968,Woosley2002}, which may occur for high-mass ($\gtrsim 140$ \Msun) Pop III stars \citep[e.g.][]{Schneider2004,Takahashi2016,deBennassuti2017}, but this would further enhance the amount of carbon due to their efficient ejection of carbon-rich material. For $\alpha$-elements such as magnesium (Mg) and calcium (Ca), Aeos20 also tends to show higher [X/Fe] ratios, consistent with the \citet{HegerWoosley2010} yields and the expectation that more massive Pop III stars are more efficient producers of $\alpha$-elements during their evolution and subsequent SN explosions.

Figure \ref{fig:p2_XH} also demonstrates an enhanced odd-even effect, where odd-Z elements such as sodium (Na) and manganese (Mn) are produced in lower relative abundances compared to their even-Z counterparts (e.g., Mg, Ca). This effect is generally more pronounced in Aeos20, as the larger number of very massive stars amplifies the odd-even disparity due to differences in nuclear reaction rates. The scatter, shown by the error bars, indicates significant halo-to-halo variation with tighter distributions for certain elements like Mn and Ca, suggesting a more uniform enrichment pattern in these elements in comparison to Fe. These results underscore how the initial mass function of Pop III stars strongly influences the chemical abundance patterns of their direct Pop II descendants, shaping both light and $\alpha$-element production as well as odd-even abundance ratios.

\begin{figure}[tb]
\center
\includegraphics[width=0.9\linewidth]{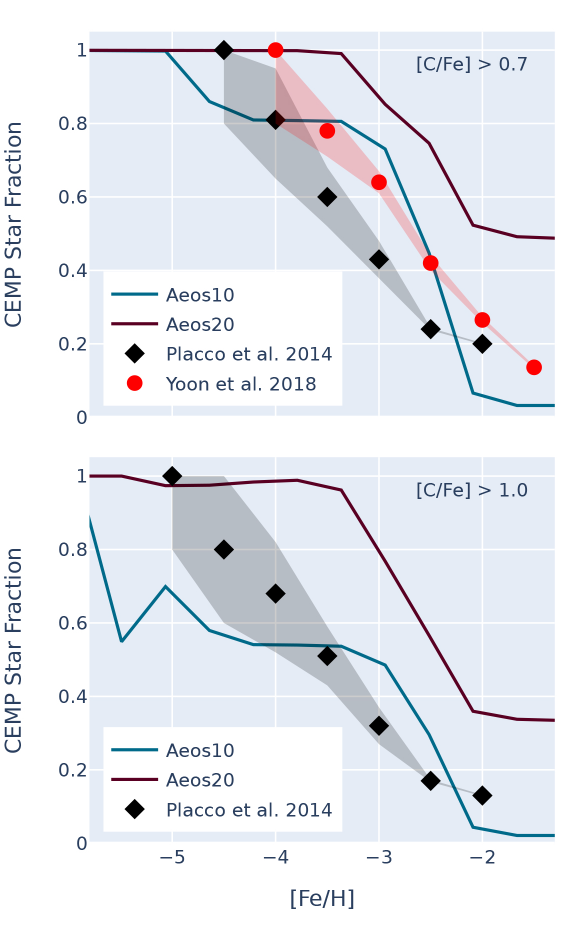}
\caption{Cumulative fractions of carbon-enhanced metal-poor (CEMP) stars as a function of metallicity. Top panel shows CEMP stars with [C/Fe] > 0.7 and bottom panel shows CEMP stars with [C/Fe] > 1.0. The low-mass Aeos10 stars are shown as a teal line, the low-mass Aeos20 stars are shown as a maroon line, observations from \citet{Placco2014} are shown as diamonds, and observations from \citet{Yoon2018} are shown as circles. The Pop III IMF significantly affects the fraction of CEMP stars, and neither of the Aeos10 nor Aeos20 IMFs reproduce Milky Way observations.
\label{fig:CEMP}}
\end{figure}

The enhanced carbon yield of more massive Pop III stars results in different amounts of carbon-enhanced metal-poor (CEMP) stars. CEMP stars, identified by their high carbon-to-iron ratios (taken either at [C/Fe] > 0.7 or [C/Fe] > 1.0), provide key insights into the formation of the first stars and galaxies \citep{Beers2005,Jeon2017}. Figure \ref{fig:CEMP} shows the cumulative fractions of our simulated CEMP stars as a function of metallicity. Because more massive Pop III stars produce a higher ratio of carbon to iron, Aeos20 has a higher fraction of CEMP stars than Aeos10. While they reproduce the qualitative trends, compared with the observed fractions of CEMP stars in the Milky Way \citep{Placco2014,Yoon2018}, neither Aeos10 nor Aeos20 consistently reproduce observations CEMP stars at a given [Fe/H]. Generally, at the lowest metallicities that are most sensitive to Pop III enrichment, Aeos10 stars have lower fractions than observations and Aeos20 stars have higher fractions. This shows that the differences in Pop III initial mass functions between Aeos10 and Aeos20 significantly affect CEMP fractions and therefore points toward an improved IMF that can better reproduce Milky Way observations.

However, at the same time, there are inconsistencies in the fractions of carbon-enhanced metal-poor stars between different Milky Way stellar surveys \citep{Arentsen2022,Arentsen2025}. Observations of dwarf galaxies also tend to find lower fractions of CEMP stars as compared to the Milky Way (e.g., \citealt{Chiti18,  Sestito2024, Ou2025}, Yelland et al. in prep).
Furthermore, Pop III SNe likely exhibited a range of explosion energies \citep{Kobayashi2006,Heger2003} that affected the metal retention of early galaxies \citep{Cooke2014,Rossi2024}. 
%This may be especially true for extremely massive (M~$\gtrsim140$~\Msun) Pop III stars that exploded as highly energetic pair-instability supernovae. 
Our current simulations do not capture the effects of different explosion energies. Additional low-metallicity observations and simulations of larger boxes with a broader range of IMF variations and Pop III SNe explosion energies to further explore these discrepancies.

\begin{figure}[bt]
\center
\includegraphics[width=0.95\linewidth]{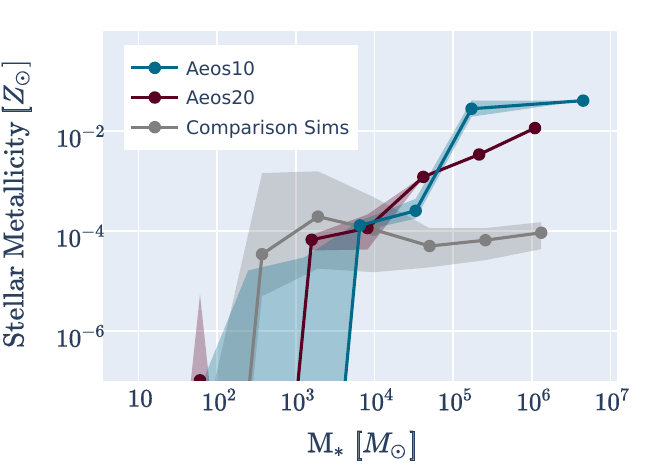}
\caption{Mass-metallicity relation for galaxies in Aeos10, Aeos20, and the comparison simulations at $z=14.5$. The shaded regions represent 16th to 84th percentile scatter.
\label{fig:stellmetalhalos}}
\end{figure}

While there are differences in the enrichment of individual metals, the differences in Pop III IMF do not produce a noticeable difference in the stellar mass-metallicity relationships of the Aeos10 and Aeos20 simulations, as seen in Figure \ref{fig:stellmetalhalos}. This may be because the increased metal yields from more massive Pop III stars in Aeos20 are counterbalanced by their lower metal ejection efficiency. As modeled by \citet{HegerWoosley2010}, massive Pop III stars often undergo fallback SNe, where a portion of the metals produced in the core region falls back onto the nascent black hole. This resulting metal retention depends on the explosion energy and stellar mass, and reduces the metals ejected into the interstellar medium and potentially limits the overall enrichment effect of these stars. Since we use the yields of \citet{HegerWoosley2010}, our results inherently reflect this mechanism. 
Additionally, many of the more massive halos in our models have begun forming Pop II stars, which quickly dominate their hosts' enrichment, obscuring differences from Pop III enrichment. To study the impact of different Pop III IMFs on chemical enrichment, it is therefore essential to focus exclusively on the oldest Pop II stars.
% can we quote a metallicity/stellar age when it becomes prohibitively difficult to see pop III features due to this? varies significantly, so unclear

In observations, the very faintest dwarf galaxies appear to exhibit a metallicity floor rather than further extending the mass-metallicity relationship found for the more luminous galaxies \citep[e.g.,][]{Ahvazi2024,Heiger2024}. Across three orders of magnitude in luminosity ($10^2 - 10^5$ \Msun), ultra-faint dwarf galaxies show a nearly constant mean metallicity, scattered around [Fe/H]$\sim$-2.6 \citep{Fu2023}. This contrasts with the expectation that less massive galaxies should have lower average metallicities. The observed plateau is not yet fully understood, but may be caused by Pop III enrichment \citep{Wheeler2019}, a ceiling on mass outflows \citep{Kravtsov2022}, or the possibility that these galaxies were once more massive and later lost stellar mass due to tidal disruption \citep[e.g.,][]{Kirby2013}. In the \Aeos simulations, we find that Pop III enrichment results in a median metallicity floor of approximately Z = $10^{-4}$ Z$_\odot$ (see also Figure 8 of \citeauthor{AeosMethods} \citeyear{AeosMethods}) with increasing stellar metallicities for galaxies of  $\sim$10$^5$ \Msun and greater. This is a slightly lower metallicity floor than seen in observed dwarf galaxies, implying either additional Pop III enrichment in observed galaxies or that an additional mechanism is necessary.

\section{Comparison to Simulations Without Individual Stellar Feedback}

\label{subsec:noindiv}

To better understand the implications of modeling individual stars and run-to-run variations, we also ran seven comparison simulations without individual Pop II star particles, without individual stellar feedback, and without detailed metal tracing. Otherwise, these simulations have the same initial conditions, cosmological parameters, and Pop III IMF as Aeos10. 

Pop III stars are represented as individual particles in all simulations. However, the comparison simulations resolve Pop II stars as cluster particles of $\sim 1000$ M$_\odot$, as in \citet{Skinner20}, while Aeos10 and Aeos20 resolve every Pop II star with $M>2$ M$_\odot$ as an individual particle. 

\begin{figure}[tb]
\center
\includegraphics[width=0.95\linewidth]{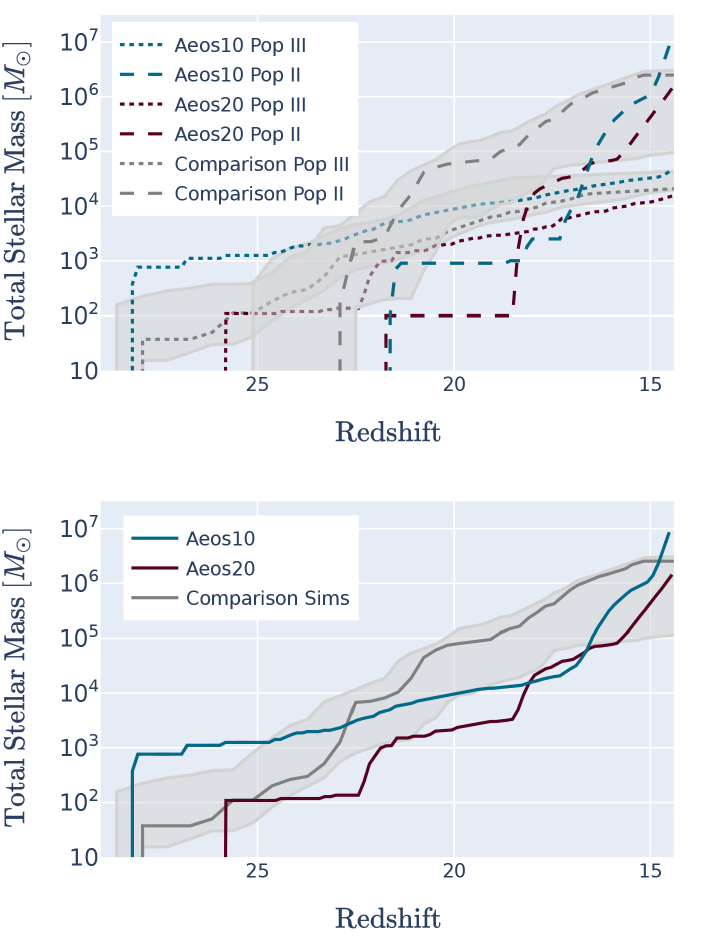}
\caption{Cumulative star formation history of the full domain down to redshift $z=14.5$. Top: Pop III and Pop II star formation of the Aeos10 volume as compared to the star formation of the comparison simulations (without individual Pop II star particles or individual feedback). The shaded gray regions show 16th to 84th percentile scatter between all comparison simulations. Bottom: Total cumulative star formation for Aeos10, Aeos20, and the comparison simulations. These bulk properties of the galaxies are not significantly affected by the inclusion of individual stars and stellar feedback.
\label{fig:fullboxsfh}}
\end{figure}

\begin{figure*}[bt]
\center
\includegraphics[width=0.9\linewidth]{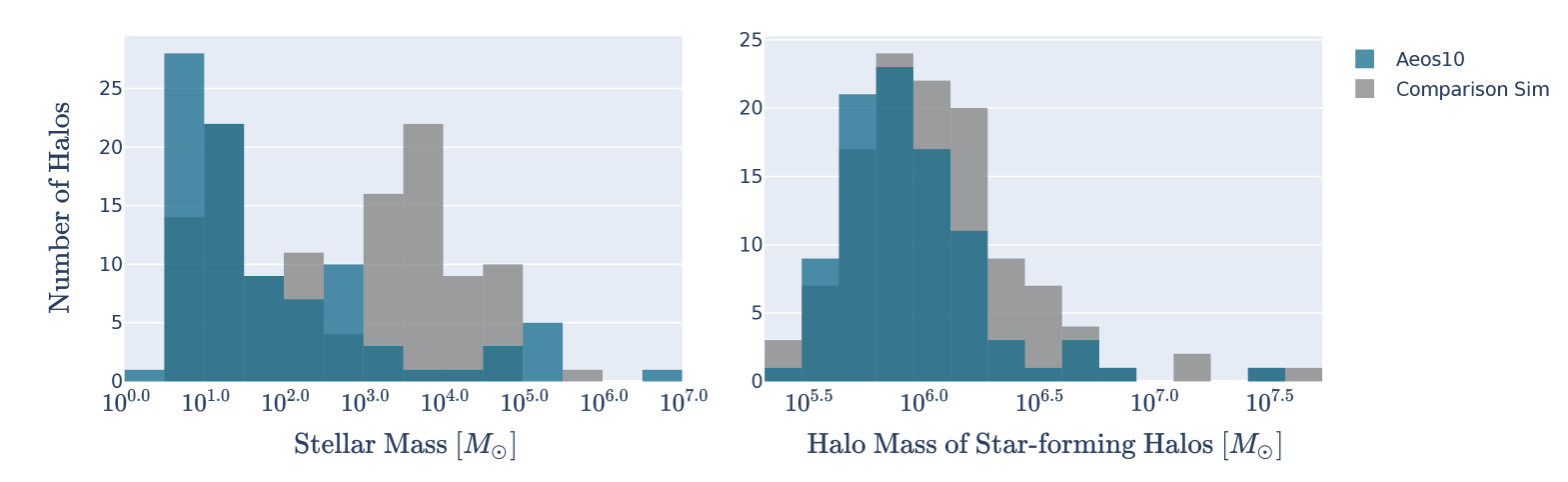}
\caption{Comparing the star-forming halos in Aeos10 with those in a typical comparison simulation. The stellar mass distributions differ; the comparison simulations generally form more galaxies around $\sim 1000$ \Msun. This is because in the comparison simulations, the Pop II stars form in particles of 1000 \Msun, biasing any halo that has begun Pop II star formation to contain at least $
1000$ \Msun. The halo mass distributions are more similar because the dark matter modeling is the same between simulations. \label{fig:haloHists_comp}}
\end{figure*}

The cumulative Pop III and Pop II star formation of the comparison simulations can be seen in the top panel of Figure~\ref{fig:fullboxsfh}. While differences exist in the overall star formation history, the differences are similar in magnitude to stochastic differences seen between identically initialized simulations, with the exception that Pop II formation tends to begin slightly earlier in the comparison simulations. 
However, the difference in onset of Pop II star formation is not large, and could be due to stochastic differences. As seen in the bottom of Figure \ref{fig:fullboxsfh}, Aeos10 and the comparison simulations do not significantly differ in their overall star formation history. In general, the bulk properties of total star formation and stellar mass-halo mass relation are not highly affected by the differences between the \Aeos simulations and the comparison simulations.

Aeos10 and Aeos20 also resemble each other, though Aeos20 starts star formation slightly behind Aeos10 ($\Delta z \simeq 3$, $\sim 15$ Myr). This arises from the interplay between the IMF and our discrete star particle spawning scheme. Both simulations adopt identical initial conditions and the same Pop~III eligibility threshold; however, once a cell first satisfies those conditions we stochastically draw individual stellar masses from the Pop~III IMF and instantiate one star particle per draw until the local gas reservoir (initialized when $\gtrsim 100\,M_\odot$ of eligible gas is present) would be overdrawn, at which point the event terminates without forming the overdrawn star. Because Aeos20 uses a more top-heavy IMF, it has a substantially higher probability that an early draw exceeds the available $\sim100\,M_\odot$ reservoir, potentially aborting the first few prospective events and forcing the halo to accumulate additional gas before a successful Pop~III burst occurs.

Figure~\ref{fig:haloHists_comp} shows the mass distribution of star-forming halos in Aeos10 vs.\ a comparison simulation. When looking only at halo mass, the distributions are similar. This appears to be because (1) the halo mass is dominated by dark matter, which does not differ between the simulations, and (2) the handling of Pop III stars is the same in the two simulations. In comparison to Figure~\ref{fig:haloHists_a20}, the masses of Pop III stars are more influential on the masses of star-forming halos than the feedback and resolution of Pop II stars.

The stellar masses, however, differ. The comparison simulations have a tendency to form galaxies of at least a few thousand solar masses because the Pop II star particles form in masses of $\sim$1000~\Msun. The number of Pop II halos in each comparison simulation differs from just a few to a few dozen, but in aggregate, there is a trend towards more galaxies around 1000~\Msun than in the \Aeos simulations. The \Aeos galaxies, on the other hand, can include Pop II galaxies with stellar masses as low as a few tens of solar masses \citep[see][]{AeosMethods}. Additionally, both the \Aeos simulations and comparison simulations can have stripped stars, resulting in the peak at a few \Msun.

\begin{figure*}[btp]
\center
\includegraphics[width=\linewidth]{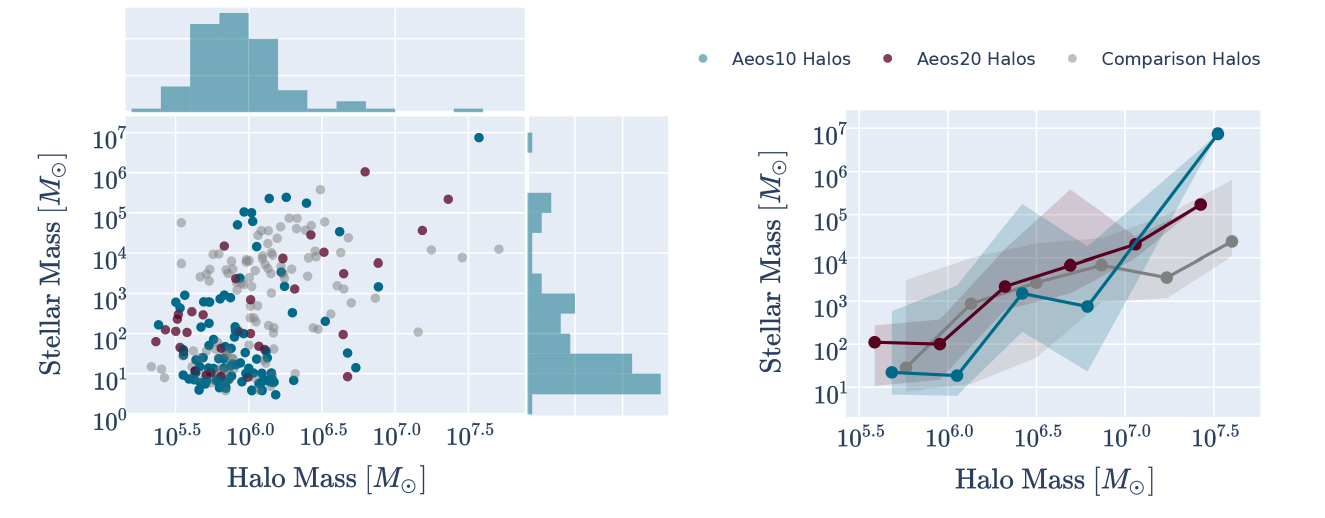}
\caption{Left: Individual stellar masses and halo masses of every star-forming halo in the simulations at redshift $z=14.5$. Shown are the halos from Aeos10, Aeos20, and a typical comparison simulation. Aeos10 is also shown in the histograms (see Figures \ref{fig:haloHists_a20} and \ref{fig:haloHists_comp} for the histograms of Aeos20 and comparison simulations). Right: Mean stellar-mass to halo-mass relation with scatter for the different simulations, summarizing the body of simulated data on the left.
\label{fig:MsMh}}
\end{figure*}

When looking at the overall $M_*-M_\text{halo}$ relations of the different simulations in Figure \ref{fig:MsMh}, however, there is not a large difference. This is largely because the scatter in the relation within a given simulation (due both to stochastic differences between comparison simulations and scatter in halos within each simulation) is just as significant as the differences between simulations. For halos greater than $\gtrsim 10^7$ \Msun, the \Aeos simulations start to form galaxies with more stellar mass than the comparison simulations. This suffers from small number statistics, though. The comparison simulations also are forced towards Pop II galaxies of at least 1000 \Msun, as previously discussed.

The metallicities of the galaxies in the comparison simulations differ more significantly from those of the \Aeos galaxies. Figure~\ref{fig:stellmetalhalos} shows the median mass-metallicity relation for each simulation with 16\% to 84\% scatter. The comparison simulations have a much shallower mass-metallicity relation than the \Aeos simulations. The \Aeos simulations have a log--log slope of $\sim 1$, which is steeper than the typical $\sim0.3$ slope of other simulated mass-metallicity relations \citep{Collins2022}, and much steeper than the practically flat relation of all the comparison simulations. The steeper \Aeos slope in contrast to the slope of the comparison simulations may be due in part to the star-by-star modeling. Both our previous work and other star-by-star simulations \citep{AeosMethods,Jeon2024,Andersson24} have found that the relatively weaker feedback of the star-by-star method (compared to burstier, stronger feedback from traditional single stellar populations) leads to the possibility of higher stellar masses at a given halo mass --  we start to see this in the stellar masses of our most massive galaxies, and likely do not see it at lower stellar masses in part due to the stellar mass resolution of the comparison simulations. This relatively weaker feedback could also allow greater retention of metals, and thus a steeper mass-metallicity relation as seen in Figure \ref{fig:stellmetalhalos}. In the comparison simulations, the aggregated bursty feedback from the more massive cluster particles (which quickly dominate galaxies with $M_* > 10^3$~\Msun) appear to eject metals more efficiently, reducing the expected increase in metallicity in halos between 10$^3$~\Msun and 10$^6$~\Msun in stellar mass. These halos are generally still low in total mass and have high loss fractions, making them highly sensitive to increases in feedback \citep[see also][]{Mead2024b}.
We note, though, that the slope at higher halo mass is based on very few galaxies and the slope at lower halo mass is dominated by scatter.
In particular, the difference in the largest galaxy and the metallicity of its surrounding galaxies between Aeos10 and the comparison galaxies is mostly due to a single burst of Pop II formation that occurs at the end of the Aeos10 run, so we refrain from drawing conclusions about $M_* > 10^5$ M$_\odot$ galaxies until longer-running simulations have been completed.

In summary, most bulk properties do not significantly differ between Aeos10 and the comparison simulations. One difference is that the comparison simulations are unable to properly form Pop II galaxies below the aggregated star particle resolution, resulting in a tendency to form more galaxies around the resolution limit. The inclusion of individual Pop II stellar feedback also tends to result in less bursty, overall weaker feedback in comparison to traditional single stellar populations, allowing for greater accumulation of metals and stellar mass.

\section{Conclusions}
\label{sec:conc}

Our analysis of the \textsc{Aeos} simulations \citep{AeosMethods} shows how varying the Pop III IMF and implementing a star-by-star feedback model influence early galaxy formation and evolution. We run two simulations with different Pop III IMFs: Aeos10, with a characteristic Pop III mass of 10 \Msun and an upper cutoff of 100 \Msun, and Aeos20, with a characteristic Pop III mass of 20 \Msun and an upper cutoff of 300 \Msun. By comparing the Aeos10 and Aeos20 simulations, we demonstrate that variations in the Pop III IMF significantly affect the ionization history, the number of early galaxies, and the chemical abundance patterns in subsequent Pop II stars. Aeos20, with its more massive Pop III stars, produces more ionizing photons, leading to substantially earlier ionization and consequently a suppression of small galaxy formation compared to what is occurring in Aeos10. 

The chemical enrichment of Aeos10 and Aeos20 differ, especially in first-generation Pop II stars. In Aeos20, the more massive Pop III stars produce yields with enhanced amounts of light elements and $\alpha$-elements. The more massive Pop III stars also produce yields with an enhanced odd-even effect, increasing the amount of elements with even atomic numbers (e.g., Mg) relative to odd-numbered elements (e.g., Na). This is reflected in the stellar chemical abundances of the first-generation Pop II stars. Aeos20 also produces a greater fraction of CEMP stars, showing that CEMP stars provide important insights into the properties of the first stars. The scatter in chemical abundances between Pop II stars from different galaxies is high though, especially in light elements, and dominates many of the differences resulting from the Pop III IMF.

We also run a suite of comparison simulations that lack individual star modeling. Comparisons with these simulations reveal that the high stellar resolution and star-by-star feedback in \Aeos create differences in the stellar mass distribution and the mass-metallicity relation. 
While Pop II star formation is still relatively new, the comparison simulations tend to form more galaxies approximately around the mass of their Pop II aggregate star particle resolution, 1000 \Msun, while the \Aeos simulations avoid this by allowing Pop II stars to form from less massive gas reservoirs. 
%The individual particle resolution also allows stars to move around and be stripped from their parent galaxies.
Additionally, the star-by-star approach tends to result in galaxies with comparatively more stellar mass, as explored by \citet{Jeon2024,Andersson24,AeosMethods}, due at least in part to less bursty feedback.
This enables a greater accumulation of metals after Pop II star formation begins and consequently steeper mass-metallicity relations compared to simulations that represent stellar populations as single particles.

Despite these differences, most of the bulk properties, such as the total star formation history, remain fairly consistent across all simulations. This highlights the robustness of certain galaxy formation trends while emphasizing the sensitivity of small-scale processes to detailed modeling choices.

Ultimately, this work underscores the importance of carefully considering the Pop III IMF and feedback resolution in simulations of early galaxy formation. As uncertainties in the Pop III IMF persist, future studies must explore the interplay of these parameters to refine our understanding of the first galaxies and their role in cosmic reionization. Further observational data on metal-poor stars and additional simulations extending to later epochs will provide constraints on these early processes.

%\acknowledgments
\section*{Acknowledgments} K.B. is supported by an NSF Astronomy and Astrophysics Postdoctoral Fellowship under award AST-2303858. J.M. acknowledges support from the NSF Graduate Research Fellowship under grant DGE-2036197. J.H.W. acknowledges support by NSF grant AST-2108020 and NASA grants 80NSSC20K0520 and 80NSSC21K1053.  M.-M.M.L. and E.P.A were partly supported by NSF grant AST-2307950 and NASA Astrophysical Theory grant 80NSSC24K0935.  G.L.B. acknowledges support from the NSF (AST-2108470 and AST-2307419, ACCESS), a NASA TCAN award, and the Simons Foundation through the Learning the Universe Collaboration.
A.P.J. acknowledges support from NSF grant AST-2307599. A.F. acknowledges support from  NSF grant AST-2307436.

The authors acknowledge the Texas Advanced Computing Center at The University of Texas at Austin for providing HPC and storage resources under project AST20007, which supported the research presented in this paper.

%************* APPENDICES ************************%

% \appendix
% \setcounter{section}{0}%
% \renewcommand\thesection{\thechapter.\Alph{section}}
% \counterwithin{figure}{section}

% \section{Appendix A}

%
% Place appendix here%

\bibliographystyle{yahapj}
\bibliography{refs}

\end{document}